\documentclass[runningheads]{llncs}
\usepackage[T1]{fontenc}
\usepackage{graphicx}
\usepackage{tabularx}
\usepackage{amsmath,amssymb}
\usepackage{subfigure}
\usepackage[dvipsnames]{xcolor}
\usepackage[norelsize,ruled,vlined,linesnumbered]{algorithm2e}
\usepackage{booktabs}
\usepackage{hyperref}
\usepackage{cleveref}

\newcommand{\vs}{\vspace{1.0mm}}

\begin{document}
\title{Simple and Fast Algorithm for Graph-based Filtered Approximate Nearest Neighbor Search (Full Version)}
\titlerunning{Simple and Fast Algorithm for Graph-based Filtered ANNS}
%
\author{
Reon Uemura\inst{1}\orcidID{0009-0004-2563-5720} \and
Keito Kido\inst{1}\orcidID{0009-0007-8550-3844} \and
Daichi Amagata\inst{1}\orcidID{0000-0001-8571-4931}
}
\authorrunning{Uemura et al.}

\institute{The University of Osaka, Osaka, Japan\\
\email{\{uemura.reon, kido.keito, amagata.daichi\}@ist.osaka-u.ac.jp}}

\maketitle              

\begin{abstract}
It has been common to represent many objects as high-dimensional vectors due to the proliferation of machine learning-based embedding techniques.
One of the most important functions for analyzing high-dimensional vectors is approximate nearest neighbor search, which, given a query vector, retrieves the vector that is approximately the most similar to the query vector.
In many real-world applications, such as e-commerce, objects have not only vectors but also attributes, e.g., category, color, and brand, and they require a scenario where users can specify a query vector and a value for each attribute of interest.
This problem, called filtered approximate nearest neighbor search, retrieves approximate nearest neighbors from a set of objects that have the specified attribute values.
Efficiently solving this problem is challenging because it has to accept arbitrary query vectors and attribute values, which are not known in advance.
Existing techniques suffer from slow search performance and difficulty in dealing with arbitrary combinations of attributes.
This work overcomes these challenges and proposes a new algorithm for this problem.
We conduct extensive experiments, and the results demonstrate the efficiency of our algorithm.
\end{abstract}
\section{Introduction}  \label{sec:introduction}
Because of the proliferation of machine learning-based embedding techniques, many objects, e.g., images and documents, are usually represented as high-dimensional vectors.
Along with this observation, vector database management systems, such as Milvus \cite{wang2021milvus}, are becoming more important.
One of the most important functions in these systems is \textit{approximate $k$ nearest neighbor search} (A$k$NNS), which, given a query vector $q$, a result size $k$, and a set $X$ of vectors, retrieves $k$ vectors $\in X$ that are approximately the most similar to $q$.
As the A$k$NNS problem finds many real-world applications, such as clustering \cite{amagata2022scalable}, EMST \cite{kido2025fast}, and RAG, existing works have studied efficient algorithms.
It is well known that obtaining the exact nearest neighbor in high dimensions is time-consuming \cite{li2019approximate}, so they focus on approximate answers.

In some real-world applications, more complex retrieval scenarios are required \cite{cai2024navigating,gollapudi2023filtered,li2025attribute,patel2024acorn,uemura2025ivf++,uemura2024efficient,wang2023efficient}.
A typical setting is attribute filtering.
Each object is associated with a vector and attributes, and a filtered A$k$NNS retrieves $k$ objects such that their vectors are approximately the most similar to $q$ among a set of objects that have the specified attribute values.
For example, in e-commerce applications, users want to search for products that are similar to their query and have a specific color and size \cite{gupta2023caps}.

\vs
\noindent
\textbf{Motivation: challenges in existing techniques.}
Standard baseline algorithms for the filtered A$k$NNS problem are Pre-filter \cite{wang2021milvus}, Post-filter \cite{yu2023pecann}, and Inline processing \cite{gollapudi2023filtered}.
Pre-filter first retrieves all objects that have the specified attribute values and then computes the $k$ nearest neighbors among them.
Although Pre-filter can obtain the exact answer, it is efficient only when the number of objects with the specified attribute values is small.
Post-filter builds a vanilla ANNS index offline, and given a query vector $q$ and $k$, it runs an A$k$NNS algorithm on the index.
If some objects in the result do not have the specified attribute values, Post-filter increases $k$ and re-runs A$k$NNS, which is repeated until $k$ objects having the attribute values are obtained.
This approach may run A$k$NNS many times, which incurs a substantial computational cost.
To alleviate this cost, Inline processing computes the distance between $q$ and $x$ only when $x$ has the specified attribute values.
Inline processing, however, still accesses many unnecessary objects, thus its improvement is limited.
Another approach to removing this concern is to build an index for each possible combination of attribute values, but it consumes an infeasible space and is impractical.

Some existing works improve the performance of filtered ANNS against the above baselines.
CAPS \cite{gupta2023caps} employs a hybrid approach of Pre-filter and Inline-processing.
Its building block is a decision tree based on the frequency of attribute values.
NHQ \cite{wang2023efficient} defines a distance between attributes and combines it with the distance between vectors.
A proximity graph is built based on this fused distance.
However, their applications are limited because they assume that queries always specify one value for \textit{every} attribute.
This assumption is usually not the case because, for example, users typically specify only size and color for the attribute constraint, although objects often have additional attributes (such as brand and price).
FilteredVamana \cite{gollapudi2023filtered} is a variant of Vamana \cite{subramanya2019diskann} that considers the attribute values of each object.
In FilteredVamana, two objects (or vectors) are connected if they are similar and share at least one attribute value.
FilteredVamana implicitly assumes that each object has a single attribute, as its original implementation suggests.
When each object has multiple attributes, the above connection approach does not guarantee that a given query can traverse a path (i.e., a set of objects) that has the specified attribute values.
Acorn \cite{patel2024acorn} and HQANN \cite{wu2022hqann} also have this drawback.

The current state-of-the-art algorithm, UNG \cite{cai2024navigating}, removes the above limitations and outperforms them.
UNG builds (i) a trie tree based on attributes and (ii) a proximity graph for each object set with the same attribute values.
Then, based on the connections in the trie tree, traversing between proximity graphs is enabled.
When the number of specified attributes is large, UNG traverses only some proximity graphs, so its search performance is good.
However, when the number of specified attributes is small, which is a typical case in practice, UNG needs to traverse many proximity graphs, degrading the search performance.

\vs
\noindent
\textbf{Contribution.}
This work overcomes the above limitations so that arbitrary filtered ANNS can be done efficiently, and we make the following contributions.
\begin{itemize}
    \item   We propose a novel framework for filtered ANNS.
            This framework allows for an arbitrary flat proximity graph (e.g., KNN graph \cite{dong2011efficient} and NSW \cite{malkov2014approximate}) as an index.
            The proximity graph in this framework is carefully built while considering attributes and can deal with any query attributes.
    \item   We conduct experiments on public datasets to evaluate our framework.
            Our experimental results show the efficiency of our algorithm.
\end{itemize}
This is a full version of \cite{uemura2026simple}.

\section{Preliminary}
Let $X$ be a set of $n$ objects, and each $x \in X$ is associated with a $d$-dimensional vector $\in \mathbb{R}^{d}$ and $m$ attribute (integer) values.
(As with existing works listed in \cite{shi2025filtered,li2025attribute}, we assume that $X$ is static, and its dynamic update (like \cite{yamashita2025should}) is beyond the scope of this paper.)
For example, when $m = 3$, the $n$ objects have attribute values regarding category, color, and size.
Note that $d = O(1)$ and $d$ is large, i.e., each vector is a high-dimensional one.
We use $x_{i}.v$ to denote the vector of $x_{i}$.
Similarly, we use $x_{i}.A$ to denote the attribute list of $x_{i}$, and $x_{i}.A[j]$ is the value of the $j$-th attribute of $x_{i}$.

For ease of explanation, we first define the exact filtered $k$ nearest neighbor search problem.

\begin{definition}
Consider $X$ and a list of query attributes $q.A = [q.A[1], ..., q.A[m]]$.
This list is called a query filter.
If $q.A[j]$ is null, the $j$-th attribute has no filter.
Then, $X(q.A)$ is defined as:
\begin{equation}
    X(q.A) = \{x  \in X \,|\, (x.A[i] = q.A[i]) \wedge (q.A[i] \neq \text{null}) \;\forall i \in [1,m] \}.
\end{equation}
That is, $X(q.A) \subseteq X$ is a set of objects $\in X$ that satisfy the query filter.
\end{definition}

\begin{definition}[Filtered $k$NNS]    \label{def:exact}
Given $X$, a query vector $q.v$, a query filter $q.A$, and a result size $k$, filtered $k$NNS returns $S^{*}$, such that $|S^{*}| = k$ and $\forall x \in S^{*}$, $\forall x' \in X(q.A)\backslash S^{*}$, $dist(x.v,q.v) \leq dist(x'.v,q.v)$, where $dist(\cdot,\cdot)$ is the Euclidean distance between two input vectors.
Ties are broken arbitrarily.
When $|X(q.A)| \leq k$, this problem returns $X(q.A)$.
\end{definition}

\noindent
\textbf{Problem definition.}
This paper considers an approximation version of Definition \ref{def:exact}.
The problem of filtered approximate $k$NNS (A$k$NNS) returns $S \subseteq X(q.A)$ and, we may have $S \neq S^{*}$.
To measure the accuracy of $S$, as with existing ANNS works, we use $recall@k$, which is defined as $recall@k = \frac{|S \cap S^{*}|}{k}$.

\section{Related Work}  \label{sec:related-work}
The exact $k$NNS algorithms typically rely on tree-based indices.
They do not function in high-dimensional spaces due to the curse of dimensionality.
This paper, hence, does not consider the exact algorithms.
The A$k$NNS problem has three main approaches: hashing, space partitioning, and proximity graphs.
It is well known that proximity graphs outperform other approaches.

\vs
\noindent
\textbf{A$k$NNS.}
In a proximity graph, each node corresponds to a vector, and two similar vectors (nodes) have an edge.
To traverse a proximity graph, \textit{greedy search} is typically employed.
From a start node, this greedy search computes the distance between a given query vector and each of the neighbor nodes.
Then, it traverses the neighbor with the smallest distance, which is repeated until no closer nodes are found.
This approach yields a more accurate result with faster time than hashing and space partitioning approaches \cite{li2019approximate}.
To shorten the search path in the search, existing works designed diverse graph indices \cite{azizi2025graph}.

In a KGraph \cite{dong2011efficient}, each vector is connected to its approximate K nearest neighbor vectors.
This graph index is built by \textsc{NNDescent} and in $O(n)$ time if K $= O(1)$ \cite{amagata2021fast,amagata2022fast}.
Navigable small world models, such as NSW \cite{malkov2014approximate} and HSNW \cite{malkov2018efficient}, have both long and short edges, so that the number of hops to $k$NNs is minimized.
Monotonic search networks \cite{dearholt1988monotonic} guarantee that any two nodes have a monotonic path, which is a path such that the distance to the target node becomes shorter as traversing to the next node in the path.
Building a monotonic search network requires $O(n^3)$ time \cite{dearholt1988monotonic}, so its approximate versions are considered in \cite{fu2021high,fu2019fast,peng2023efficient}.
Relative neighborhood graphs \cite{toussaint1980relative} prune redundant edges, making a proximity graph sparser and alleviating unnecessary node traversals.
Vamana \cite{subramanya2019diskann} approximates the relative neighborhood graphs.

\vs
\noindent
\textbf{Filtered A$k$NNS.}
Existing works consider matching filters \cite{cai2024navigating,gollapudi2023filtered,gupta2023caps,wang2023efficient,wu2022hqann} (i.e., objects have the same attribute values as the query filters) and range filters \cite{zuo2024serf,engels2024approximate,jiang2025digra,xu2024irangegraph} (i.e., objects have an attribute value falling in the specified range).
This paper focuses on the former case.

NHQ \cite{wang2023efficient} and HQANN \cite{wu2022hqann} fuse the distance based on vector similarity and the one based on attribute similarity.
ACORN \cite{patel2024acorn} is essentially HNSW and employs Inline-processing.
CAPS employs an IVF structure, and its clusters are built by a decision tree based on attribute values.
UNG partitions $X$ into disjoint subsets, and each subset has the same attribute values.
A label navigation graph (LNG) is built based on a trie structure and attribute containment relationships.
At the same time, for each subset, UNG builds a proximity graph.
Then, based on LNG and a given query filter, UNG traverses necessary proximity graphs.
According to the latest experimental study \cite{shi2025filtered}, UNG clearly outperforms the others, e.g., by at least an order of magnitude in QPS under the same recall, in our filter setting.
We thereby use UNG as a baseline in our experiments.
One important practical observation is that, as benchmark papers \cite{li2025attribute,shi2025filtered,zhu2025experimental} suggest, \textit{the number of attributes specified by queries is usually one or two in practice}.
UNG does not care about this observation, and in such cases, it may need to traverse the graphs of many subsets, thereby degrading search performance.

\section{Proposed Algorithm}
As mentioned earlier, proximity graphs have a better search-recall tradeoff than the other approaches.
We therefore design a new proximity graph that considers attribute filters.

\vs
\noindent
\textbf{Main idea.}
Let us consider the following three cases:
\begin{description}
    \item[Case 1.]  $|q.A| = m$, i.e., the query filter specifies all attributes.
    \item[Case 2.]  $|q.A| = 1$, i.e., the query filter specifies one of $m$ attributes.
    \item[Case 3.]  $2 \leq |q.A| < m$, i.e., the other cases.
\end{description}

\noindent
\textbf{Case 1.}
An intuitive approach to deal with case 1 is as follows.
We partition $X$ into disjoint subsets, such that the objects in a subset have the same attribute value for every attribute.
Then, we build a proximity graph for each subset.
Given a query vector $q.v$ and a query filter $q.A$, we can use the proximity graph of $q.A$ and run the greedy search on the graph.
This approach is simple yet efficient.
Furthermore, the total number of nodes is $n$, and if the number of edges is $O(1)$, the space complexity is $O(n)$, scaling well to large datasets.

\vs
\noindent
\textbf{Case 2} can also employ a similar idea to the one above.
Specifically, for the $i$-th attribute, we partition $X$ into disjoint subsets based on the attribute values and build a proximity graph for each subset.
This is applied for each $i \in [1,m]$.
Because each object is involved in $m$ proximity graphs, this approach consumes $O(nm)$ space.
As $m$ is practically small (e.g., $\leq 3$) \cite{lin2025survey}, this approach is feasible.

\vs
\noindent
\textbf{Case 3.}
For any query filter in case 1 or 2, the above approaches can find the graph index containing only objects that match the query filter.
To achieve this for case 2, we have to enumerate all possible combinations of attribute values.
This enumeration requires $O(\lambda^m)$ proximity graphs, where $\lambda$ is the average number of distinct values in each attribute, and this is infeasible in practice.

To address this issue, it is important to notice that case 3 is a subset (superset) of case 1 (2).
Let $X_{a,b,...,c}$ be a set of objects such that their 1st, 2nd, ..., and $m$-th attribute values are respectively, $a$, $b$, ..., and $c$.
Similarly, let $G_{a,b,...,c}$ be the graph index of $X_{a,b,...,c}$.
Now assume that we are given a query filter $q.A = [q.A[1] = a$, $q.A[2] = b$, ..., $q.A[m] = null]$.
It is easy to see that all objects in $X_{a,b,...,c}$ satisfy $q.A$, so $G_{a,b,...,c}$ is available.
However, the objects in $X_{a,b,...,*}$, where $*$ is a wildcard, also have to be considered, so using only $G_{a,b,...,c}$ may result in low recall.

Our idea that overcomes this challenge: If there exist edges between $G_{a,b,...,c}$ and $G_{a,b,...,*}$, we can traverse only nodes (objects or vectors) that satisfy $q.A$ and are closer to $q.v$.
This idea suggests that we do not have to traverse every $G_{a,b,...,*}$ for $q.A = [q.A[1] = a$, $q.A[2] = b$, ..., $q.A[m] = null]$.
To create such edges while minimizing space consumption, we add the edges made in case 2 to the graphs built in case 1.
Let $\gamma$ be the average degree of the graph indices in case 2, and each object eventually has $O(m\gamma)$ edges on average.
Note that some neighbors do not satisfy a given query filter, and we do not need to traverse them like Inline-processing.

\subsection{Index Structure}
\textbf{Overview.}
Our new graph $\mathcal{G} = \langle X,E\rangle$ is built by merging multiple graphs.
Let $G_{a,b,...,c} = \langle X_{a,b,...,c}, E_{a,b,...,c} \rangle$, where $a$, $b$, and $c$ are variable, whereas $E_{a,b,...,c}$ is a set of directed edges with labels.
(We later define labels.)
We build $G_{a,b,...,c}$ for all $X_{a,b,...,c} \subset X$.
Similarly, we build $G_{a,*,...,*}$, $G_{*,b,...,*}$, ..., and $G_{*,*,...,c}$ for all $X_{a,*,...,*}$, $X_{*,b,...,*}$, ..., and $X_{*,*,...,c} \subset X$.
Each object has edges generated in these graphs.

\vs
\noindent
\textbf{Edge label.}
Consider two objects in $\mathcal{G}$, and let $l(x,x')$ be the label of the directed edge $(x,x') \in E$, i.e., $x \rightarrow x'$.
It is defined as follows.
\begin{equation}
    l(x,x') = 
    \begin{cases}
        \perp   & (\text{if}\; x.A = x'.A)   \\
        x'.A & (\text{otherwise})
    \end{cases}
    \label{eq:label}
\end{equation}

\begin{figure}
    \centering
    \includegraphics[width=0.6\linewidth]{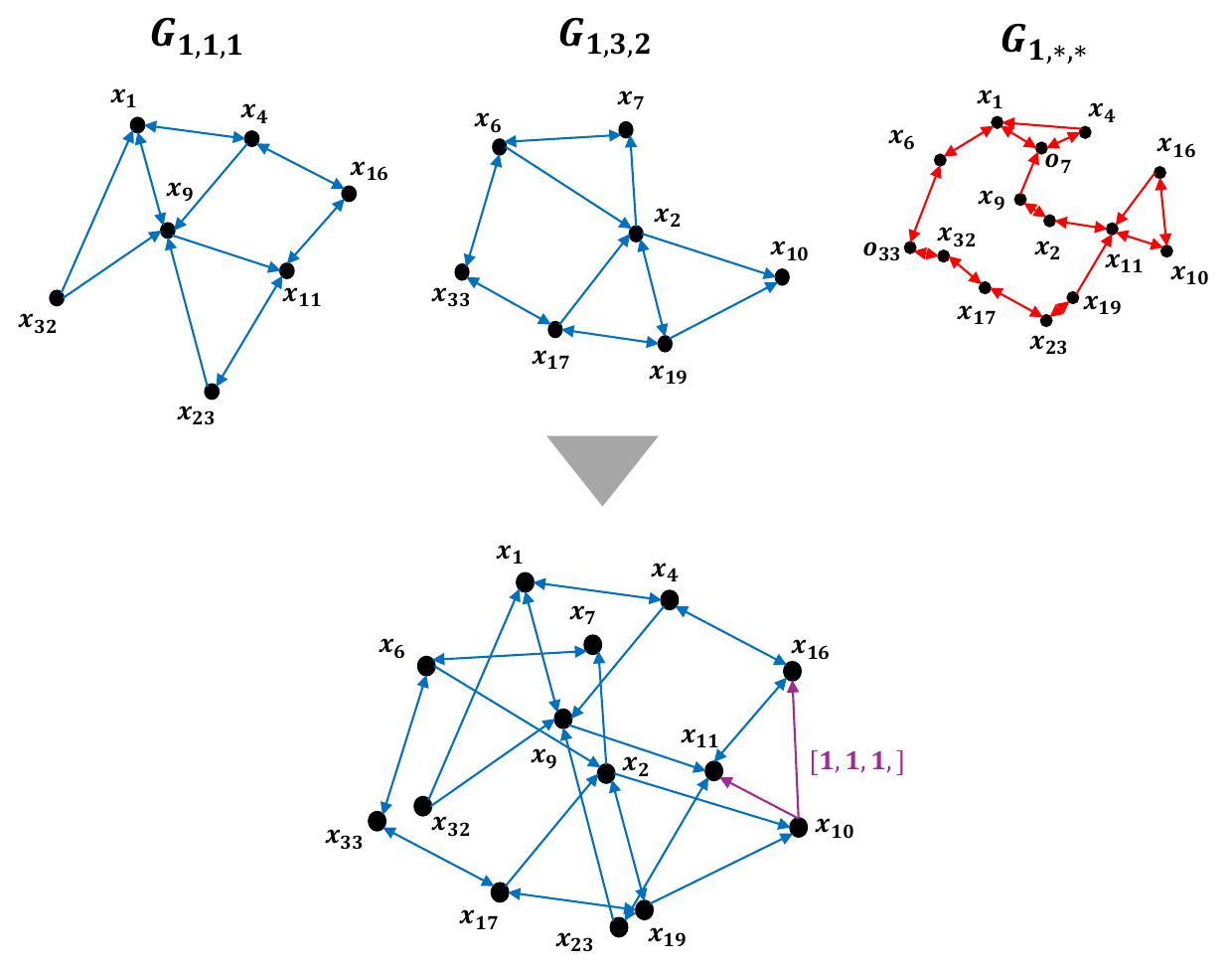}
    \caption{Example of our graph building}
    \label{fig:graph}
\end{figure}

\begin{example}
Assume that $m = 3$ and we have $X_{1,1,1}$ and $X_{1,3,2}$.
Then, $X_{1,*,*}$ contains all objects in $X_{1,1,1}$ and $X_{1,3,2}$.
The top part of \Cref{fig:graph} illustrates the graph indices of $X_{1,1,1}$, $X_{1,3,2}$, and $X_{1,*,*}$.
We add edges of $G_{1,*,*}$ to $G_{1,1,1}$ and $G_{1,3,2}$.
The bottom part of \Cref{fig:graph} shows this edge addition.
(For conciseness, not all are shown.)
The labels of the blue edges are $\perp$, whereas the label of edge $(x_{10},x_{16})$ is $[1,1,1]$ because $x_{16} \in X_{1,1,1}$.
\end{example}

\vs
\noindent
\textbf{Which graph index is available in our algorithm?}
As seen in \Cref{fig:graph}, our graph index should be \textit{flat} because of our edge addition strategy.
An arbitrary flat graph index is available in our algorithm.
In our experiments, we use NSW (with optimization in \cite{malkov2018efficient}) as our base graph index.

\vs
\noindent
\textbf{Pre-processing algorithm.}
We build our graph index with \Cref{algo:preprocessing}.
This pre-processing is done only once, since our index is available for arbitrary queries.

To start with, for each $X_{a,b,...,c} \subseteq X$, we build a proximity graph $G_{a,b,...,c}$.
In \Cref{algo:preprocessing}, \textsc{Build-Graph}$(\cdot,\cdot)$ is the building algorithm of a given graph index structure.
Since this graph contains only objects that share the same attribute values, the labels of its edges are $\perp$.

Next, for each $x \in X$, we assign it to $m$ object sets.
Specifically, let $X_{\langle i, \alpha \rangle} = \{x \in X : x.A[i] = \alpha \}$, and for each $i \in [1,m]$, we add $x$ into $X_{\langle i, x.A[i]\rangle}$.
After we do this for every $x \in X$, we build a proximity graph for each $X_{\langle i, x.A[i]\rangle}$ and add its edges into $\mathcal{G}$ while making their labels based on \Cref{eq:label}.
Notice that $\mathcal{G}$ is considered to be a set of $G_{a,b,...,c}$ with augmented edges from $G_{a,*,...,*}$, $G_{*,b,...,*}$, ..., and $G_{*,*,...,c}$.

\begin{algorithm}[!t]
    \caption{\textsc{Pre-Processing}}
    \label{algo:preprocessing}
    \DontPrintSemicolon
    \KwIn {$X$, $m$, $K$ (initial graph degree), $K'$ (hyper-parameter)}
    $\mathcal{G} \gets \varnothing$\;
    \ForEach {$X_{a,b,...,c} \subseteq X$}{
        $G_{a,b,...,c} \gets$ \textsc{Build-Graph}$(X_{a,b,...,c}, K)$\;
        $\mathcal{G} \gets \mathcal{G} \cup G_{a,b,...,c}$ // $\forall e \in E, l(e) = \perp$
    }
    \ForEach{$x \in X$}
    {
        \ForEach{$i \in [1,m]$}
        {
            $X_{\langle i, x.A[i]\rangle} \gets X_{\langle i, x.A[i]\rangle} \cup \{x\}$
        }
    }
    \ForEach{$X_{\langle i, x.A[i]\rangle}$}
    {
        $G' \gets$ \textsc{Build-Graph}$(X_{\langle i, x.A[i]\rangle}, K')$\;
        \ForEach{$(x,x') \in E'$ such that $(x,x') \notin E$}
        {
            $E \gets E \cup \{(x,x')\}$ with $l(x,x') = x'.A$
        }
    }
    \textbf{return} $\mathcal{G}$
\end{algorithm}

\vs
\noindent
\textbf{Time complexity.}
Let $\mathcal{T}(n)$ be the time of \textsc{Build-Graph} of $n$ objects.
Assuming that the graph degree is $O(1)$, the first for loop in \Cref{algo:preprocessing} requires $\sum\mathcal{T}(|X_{a,b,...,c}|) = \mathcal{T}(n)$, since $\bigcup X_{a,b,...,c} = X$ and $X_{a,b,...,c}$ is disjoint.
The second for loop trivially requires ${\rm\Theta}(nm)$, as $\sum|X_{\langle i, x.A[i]\rangle}| = nm$.
The last for loop involves $nm$ objects, so $\mathcal{T}(nm)$ time is required.
To summarize, the time complexity of \Cref{algo:preprocessing} is $\mathcal{T}(nm)$.

\vs
\noindent
\textbf{Space complexity.}
As our index requires a flat graph structure, the number of nodes is $n$.
Since each graph built in the first and last for loops has a constant degree, the number of edges is $O(nm)$.
Therefore, the space complexity of our index is $O(nm)$.

\subsection{Query Processing}
Now we are ready to present our proposed algorithm, which is described in \Cref{algo:anns}.
Note that $\epsilon$ is a parameter that controls the tradeoff between search time and recall.

\begin{algorithm}[t]
    \caption{\textsc{Filtered ANNS}}	\label{algo:anns}
    \DontPrintSemicolon
    \KwIn{$\mathcal{G}$, $q.v$, $q.A$, $\epsilon$ ($\geq 1$), and $k$}    
    \tcc{Determining a start node}
    $A \gets q.A$\;
    \For{all $i \in [1,m]$ such that $q.A[i] = null$}
    {
        Replace $q.A[i]$ with a random value in the domain of the $i$-th attribute 
    }
    $x_s \gets$ the medoid node in $G_{A}$\;
    \tcc{Traversing the graph from the start node}
    $V \gets \{x_{s}\}$ // $V$ is a set of visited nodes\;
    $P, P' \gets \langle x_{s}, dist(x_{s}.v, q.v) \rangle$ // $P$ and $P'$ are sorted by $dist(\cdot, q.v)$)\;
    $\tau \gets \infty$\;
    \While{$P' \neq \emptyset$}
    {
        $x \leftarrow $ the top element in $P'$\;
        Pop $P'$\;
        \ForEach {$x'$ such that $(x,x') \in E$ and $x' \notin V$}
        {
            $V \gets V \cup \{x'\}$\;
            \eIf {$l(x,x') = \perp$}
            {
                \If {$dist(x'.v,q.v) < \tau$}{  \label{algo:anns:update-queue_b}
                    Add $\langle x', dist(x'.v, q.v) \rangle$ into $P$ and $P'$\;
                    Update $\tau$ // the $\epsilon \cdot k$-th $dist(\cdot,q.v)$ in $P$\;
                    \If {$|P| > \epsilon \cdot k$}
                    {
                        Erase the last element in $P$ \label{algo:anns:update-queue_e}
                    }
                }
            }
            {
                \If {$x'.A$ satisfies $q.A$}
                {
                    Run lines \ref{algo:anns:update-queue_b}--\ref{algo:anns:update-queue_e}
                }
            }
        }
    }
    \textbf{return} the first $k$ elements in $P$
\end{algorithm}

First, we determine the start node $x_s$ from $\mathcal{G}$.
If $q.A$ has no null, $x_s$ is set as the medoid node of $G_{q.A}$ (which is computed in pre-processing).
On the other hand, if $q.A$ has a null, we assign a random value in the domain for each null attribute in $q.A$.
Then, $x_s$ is determined in the same way as the case where $q.A$ has no null.

Next, we run the greedy search from $x_s$.
In a nutshell, we traverse unvisited neighbors, which are then marked as visited and pushed into a priority queue based on the distance to $q.v$.
Note that we visit only neighbors such that the labels of the corresponding edges are $\perp$ or satisfy $q.A$.
We repeat these operations until no closer objects are found.

\section{Experiment}
This section reports our experimental results.
All experiments were conducted on a Ubuntu 24.04 LTS machine with an Intel Xeon Gold 6254@3.10GHz CPU, and 768GB RAM.

\vs
\noindent
\textbf{Dataset.}
We used SIFT, Deep, and GIST, which are commonly used in filtered ANNS works \cite{gollapudi2023filtered,gupta2023caps}.
SIFT and Deep have 100 million vectors, while GIST has 1 million vectors.
By default, we used 1 and 10 million vectors for SIFT and Deep, respectively. 
(The full set is used in our scalability test.)
We set $m = 3$, as in \cite{cai2024navigating,gupta2023caps,wang2022navigable}, so each object has three attributes.
For each attribute, we employed the power law distribution, as used in \cite{cai2024navigating}, because it is commonly observed in real-world settings \cite{gupta2023caps}.
The first, second, and third attributes have 68, 10, and 12 distinct values, which is a more challenging setting than that in \cite{cai2024navigating}.

For each dataset, we used 1,000 queries obtained from the corresponding source.
Their query filters also followed the same power law distribution.
When a query filter specified a single attribute, a random attribute was chosen.
Similarly, when a query filter specified two attributes, a random single attribute had a null value.
We ran each query in single-thread mode.

\vs
\noindent
\textbf{Algorithm.}
We compared our algorithm with UNG \cite{cai2024navigating}, a state-of-the-art filtered ANNS algorithm.
We did not consider the other algorithms, because state-of-the-art benchmarking works \cite{cai2024navigating,li2025attribute,shi2025filtered,zhu2025experimental} demonstrate that UNG clearly outperforms them.
For UNG, we used the original implementation\footnote{\url{https://github.com/YZ-Cai/Unified-Navigating-Graph}} with the default parameters.
For our algorithm, $K = K' = 16$ (see Algorithm \ref{algo:preprocessing}).
They were implemented in C++ and compiled by g++ with -O3 optimization.
SIMD instruction was enabled.

\begin{figure*}[!t]
    \begin{center}
        \subfigure[SIFT]{%
            \includegraphics[width=0.31\linewidth]{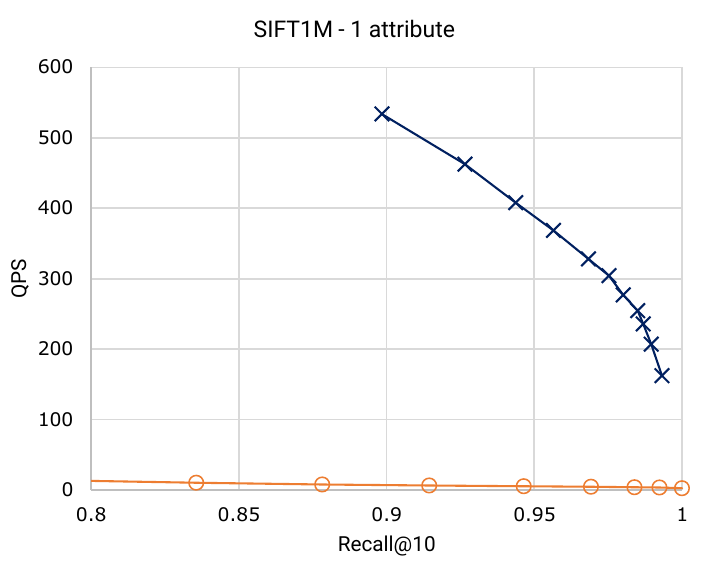}   \label{fig:sift_1}}
        \subfigure[GIST]{%
            \includegraphics[width=0.31\linewidth]{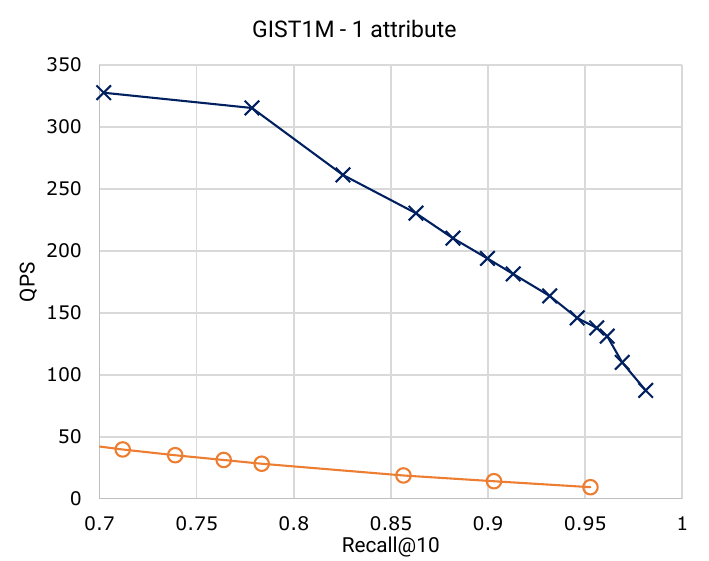}   \label{fig:gist_1}}
        \subfigure[Deep]{%
            \includegraphics[width=0.31\linewidth]{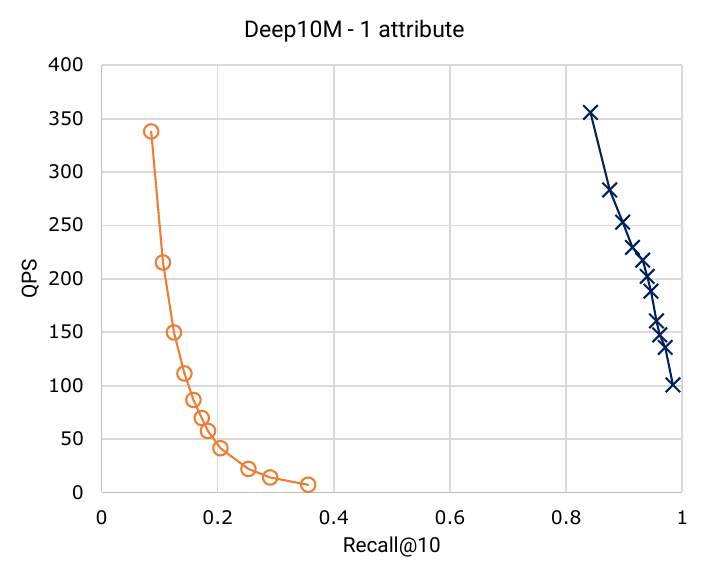}   \label{fig:deep_1}}
        \caption{QPS vs. recall@10 for query filters with a single attribute.
                ``\textcolor{Blue}{$\times$}'' shows \textcolor{Blue}{\textsf{Ours}} and ``\textcolor{orange}{$\circ$}'' shows \textcolor{orange}{\textsf{UNG}}.}
        \label{fig:tradeoff_1}
    \end{center}
\end{figure*}
\begin{figure*}[!t]
    \begin{center}
        \subfigure[SIFT]{%
            \includegraphics[width=0.31\linewidth]{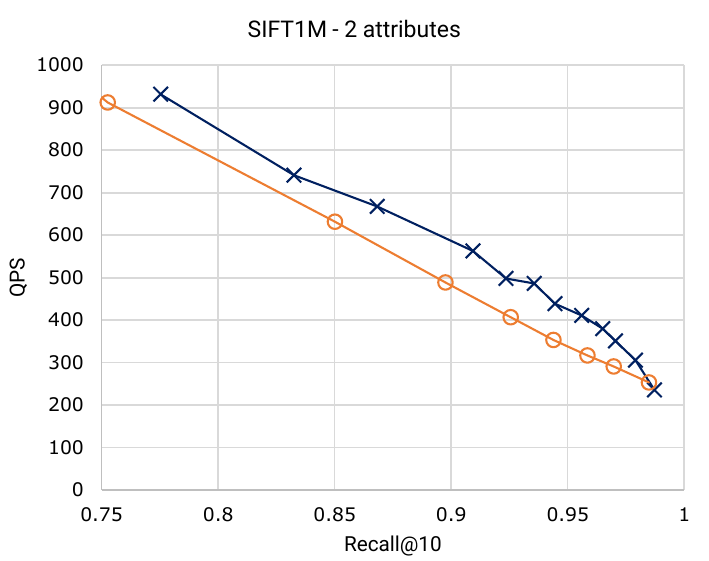}   \label{fig:sift_2}}
        \subfigure[GIST]{%
            \includegraphics[width=0.31\linewidth]{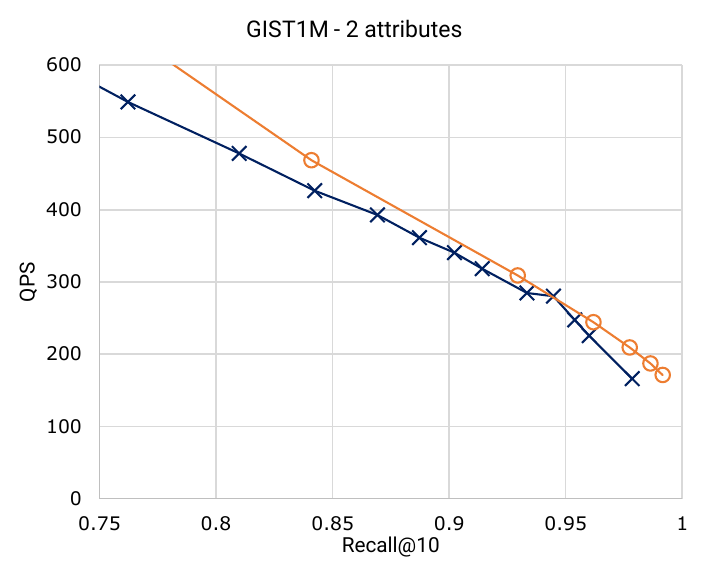}   \label{fig:gist_2}}
        \subfigure[Deep]{%
            \includegraphics[width=0.31\linewidth]{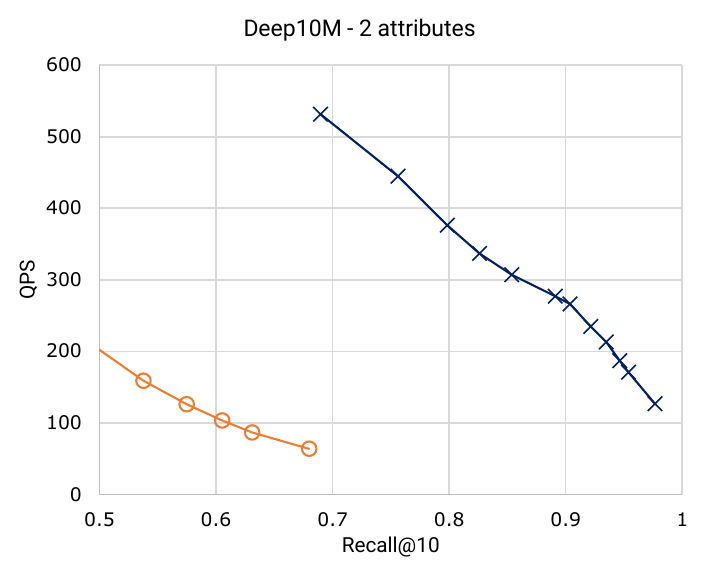}   \label{fig:deep_2}}
        \caption{QPS vs. recall@10 for query filters with two attributes.
            ``\textcolor{Blue}{$\times$}'' shows \textcolor{Blue}{\textsf{Ours}} and ``\textcolor{orange}{$\circ$}'' shows \textcolor{orange}{\textsf{UNG}}.}
        \label{fig:tradeoff_2}
    \end{center}
\end{figure*}
\begin{figure*}[!t]
    \begin{center}
        \subfigure[SIFT]{%
            \includegraphics[width=0.31\linewidth]{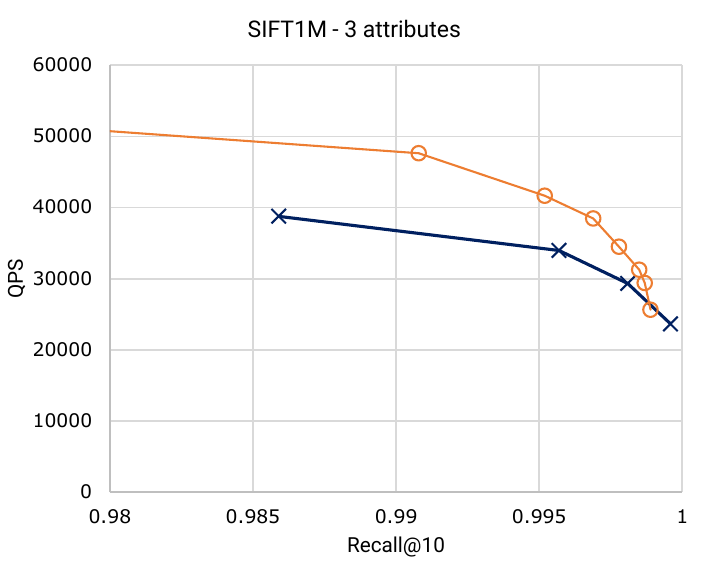}   \label{fig:sift_3}}
        \subfigure[GIST]{%
            \includegraphics[width=0.31\linewidth]{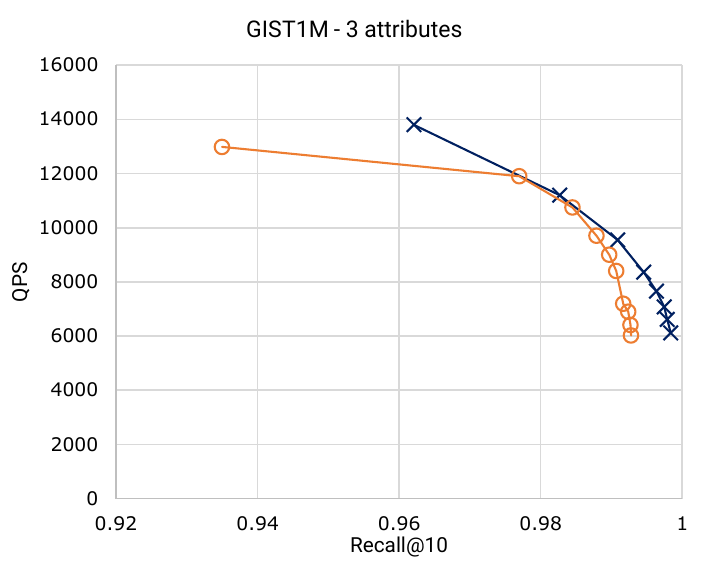}   \label{fig:gist_3}}
        \subfigure[Deep]{%
            \includegraphics[width=0.31\linewidth]{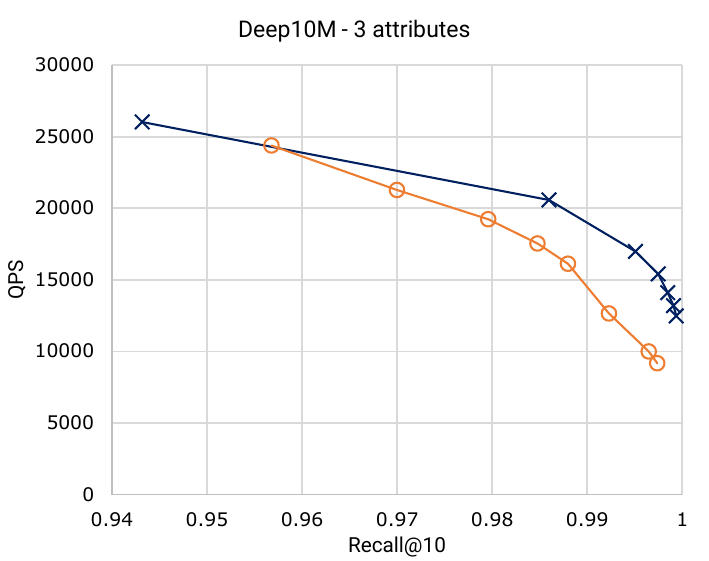}   \label{fig:deep_3}}
        \caption{QPS vs. recall@10 for query filters with three attributes.
            ``\textcolor{Blue}{$\times$}'' shows \textcolor{Blue}{\textsf{Ours}} and ``\textcolor{orange}{$\circ$}'' shows \textcolor{orange}{\textsf{UNG}}.}
        \label{fig:tradeoff_3}
    \end{center}
\end{figure*}

\vs
\noindent
\textbf{Criteria.}
We measured the A$k$NNS performance tradeoff between QPS (query per second) and recall by varying $\epsilon$.
We set $k = 10$, as with existing works.

\vs
\noindent
\textbf{Result.}
Figures \ref{fig:tradeoff_1}, \ref{fig:tradeoff_2}, and \ref{fig:tradeoff_3} exhibit the experimental results.

\vs
\noindent
\textit{Query filter of a single attribute case.}
Figure \ref{fig:tradeoff_1} shows the results when query filters have a single attribute.
They demonstrate that our algorithm substantially outperforms UNG.
In our graph index, each node has edges obtained from $G_{a,*,*}$, $G_{*,b,*}$, and $G_{*,*,c}$, so each node (object) has edges to similar objects ``w.r.t. a given query filter.''
On the other hand, UNG needs to traverse nodes based on the trie tree, i.e., an attribute-based relationship, which does not guarantee connecting similar objects.
This property significantly degrades the ANNS performance when only a subset of $m$ attributes is specified as a query filter.

\vs
\noindent
\textit{Query filter of two attributes case.}
\Cref{fig:tradeoff_2} describes the results when query filters have two attributes.
Our algorithm consistently outperforms UNG on SIFT and Deep, whereas their performances are comparable on GIST.
The effectiveness of our main idea is demonstrated in these results, as our algorithms do not build any proximity graphs for the case of two attributes.
As mentioned in Section \ref{sec:related-work}, the number of attributes specified by queries is usually one or two in practice.
Our algorithm normally outperforms the state-of-the-art in this practical scenario.

\vs
\noindent
\textit{Query filter of three attributes case.}
Figure \ref{fig:tradeoff_3} shows the results when query filters have three (i.e., all) attributes.
UNG yields a result comparable to ours, as UNG traverses only the graph consisting of objects that match a given query filter when $|q.A| = m$.
Ours also has this property, so they share a similar result (although ours keeps outperforming UNG on Deep, a larger dataset than SIFT and GIST).

\vs
\noindent
\textit{Memory consumption.}
\Cref{tab:memory} shows the index sizes of ours and UNG.
Although ours needs more space, the difference is negligible.

\begin{table}[!t]
    \centering
    \caption{Index size [GB]}
    \label{tab:memory}
    \begin{tabular}{ccc}        \toprule
        Dataset & Ours  & UNG   \\ \midrule
        SIFT    & 0.83  & 0.59  \\
        GIST    & 3.85  & 3.68  \\
        Deep    & 7.21  & 4.96  \\
        \bottomrule
    \end{tabular}
\end{table}

\vs
\noindent
\textit{Scalability test.}
Last, we investigate the scalability of our algorithm by using 1, 10, and 100 million vectors.
When scaling to 100M vectors, we were unable to run UNG successfully using the authors' public implementation under our experimental setup.
(It terminated with a runtime failure, i.e., a segmentation fault.)
Hence, we only report the scalability results of our algorithm.

\begin{figure*}[!t]
    \begin{center}
        \subfigure[1 attribute]{%
            \includegraphics[width=0.31\linewidth]{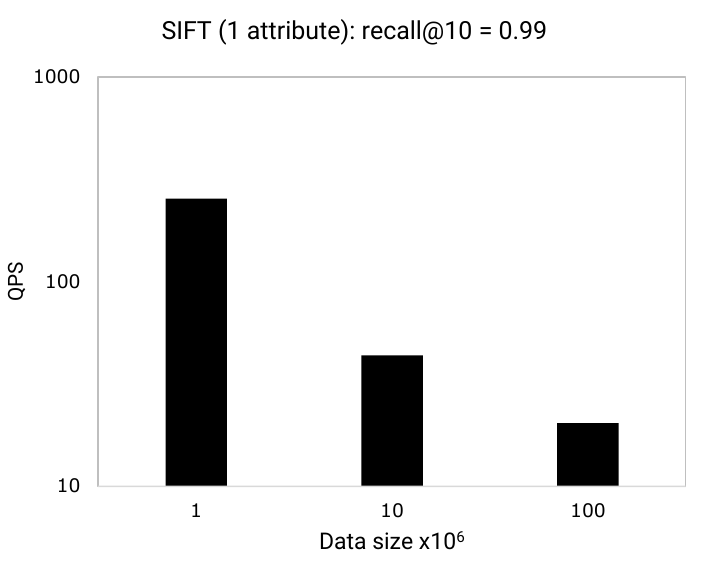}   \label{fig:sift_n_1}}
        \subfigure[2 attributes]{%
            \includegraphics[width=0.31\linewidth]{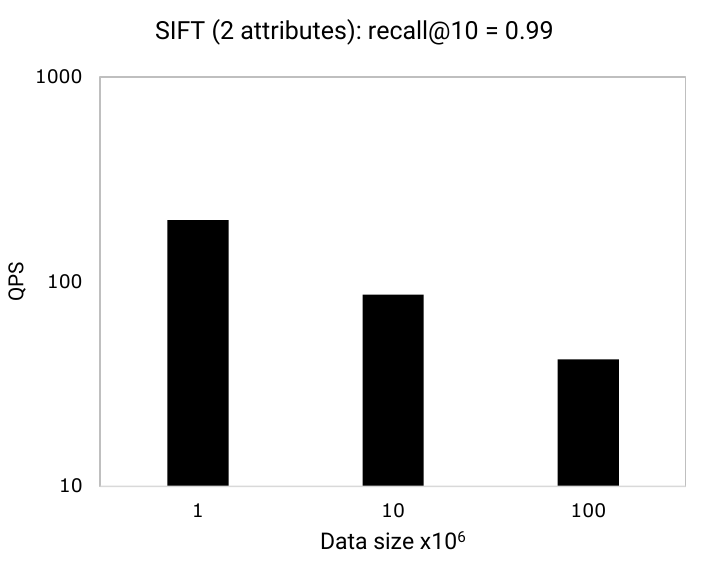}   \label{fig:sift_n_2}}
        \subfigure[3 attributes]{%
            \includegraphics[width=0.31\linewidth]{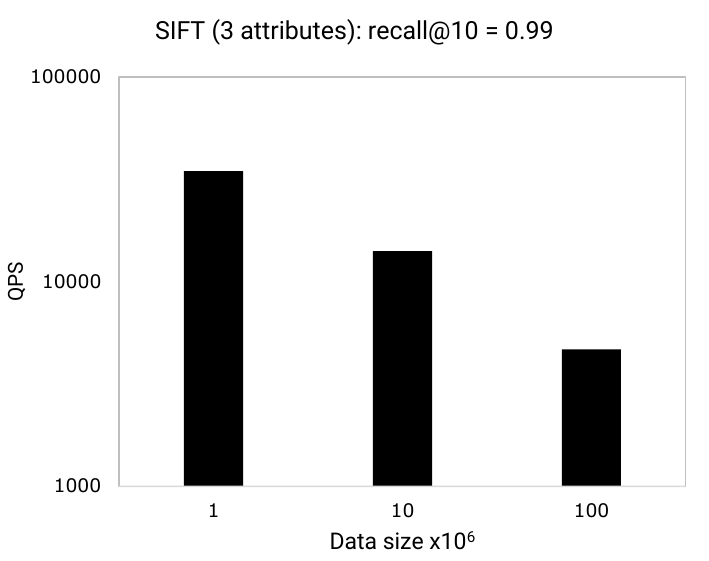}   \label{fig:sift_n_3}}
        \caption{Impact of $n$ on SIFT under $recall@10 = 0.99$}
        \label{fig:sift_n}
    \end{center}
\end{figure*}
\begin{figure*}[!t]
    \begin{center}
        \subfigure[1 attribute]{%
            \includegraphics[width=0.31\linewidth]{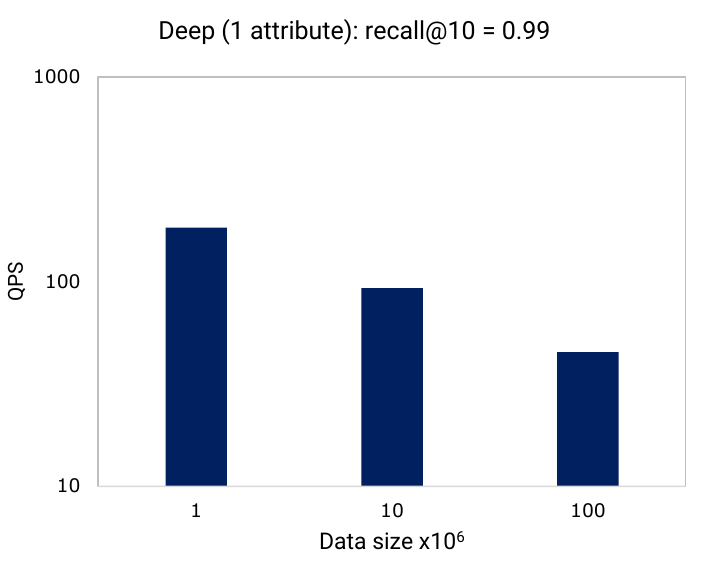}   \label{fig:deep_n_1}}
        \subfigure[2 attributes]{%
            \includegraphics[width=0.31\linewidth]{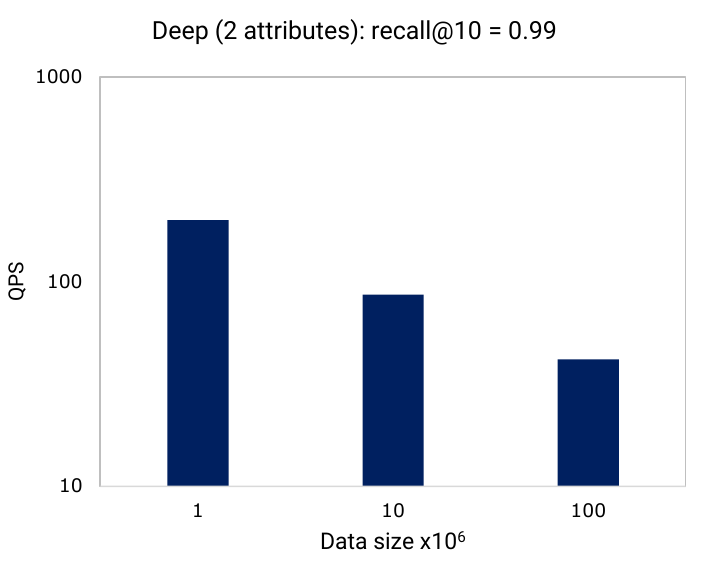}   \label{fig:deep_n_2}}
        \subfigure[3 attributes]{%
            \includegraphics[width=0.31\linewidth]{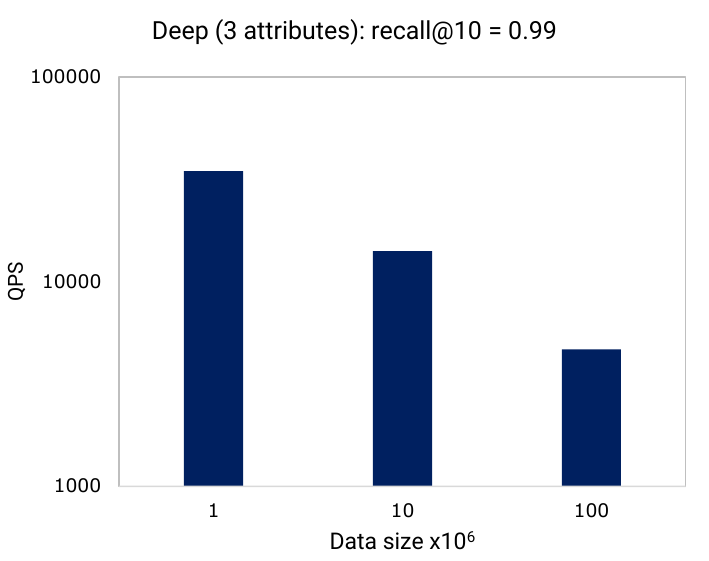}   \label{fig:deep_n_3}}
        \caption{Impact of $n$ on Deep under $recall@10 = 0.99$}
        \label{fig:deep_n}
    \end{center}
\end{figure*}

Figure \ref{fig:sift_n} (Figure \ref{fig:deep_n}) shows the QPS result on SIFT (Deep) under $recall@10 = 0.99$.
As the dataset size, i.e., $n$, grows, QPS drops, which is quite a natural phenomenon.
One important observation here is that QPS does not drop linearly with $n$ but rather \textit{sublinearly}.
This result demonstrates that our algorithm scales well to $n$.
As a reference, we note that the QPS of Pre-filter on SIFT 100M with a single-attribute query filter is 0.42.
Ours achieves more than 100 times better QPS with $recall@10 = 0.99$ in the same setting.

\section{Conclusion}    \label{sec:conclusion}
This paper tackled the problem of filtered approximate $k$ nearest neighbor search on high-dimensional data.
Existing works do not fully address this problem and have critical drawbacks, such as handling only a single attribute, not supporting the specification of attribute subsets, and inefficiency in practical use cases.
To remove these drawbacks, we proposed a new graph-based A$k$NNS algorithm.
Our experimental results show that our algorithm usually outperforms the existing state-of-the-art UNG.

\bibliographystyle{splncs04}
\bibliography{ijcai25}

\end{document}